# Spin-orbit-driven quarter semimetals in rhombohedral graphene

Jing Ding[1,2,*], Hanxiao Xiang[1,2,*], Naitian Liu[1,2], Wenqiang Zhou[1,2], Xinjie Fang[1,2], Zhangyuan Chen[1,2], Le Zhang[1,2], Kenji Watanabe[3], Takashi Taniguchi[4], Shuigang Xu[1,2,†]

[1] *Key Laboratory for Quantum Materials of Zhejiang Province, Department of Physics, School of Science, Westlake University, Hangzhou, China*
[2] *Institute of Natural Sciences, Westlake Institute for Advanced Study, Hangzhou, China*
[3] *Research Center for Electronic and Optical Materials, National Institute for Materials Science, Tsukuba, Japan*
[4] *Research Center for Materials Nanoarchitectonics, National Institute for Materials Science, Tsukuba, Japan*

[*] These authors contributed equally to this work.
[†] Corresponding author: xushuigang@westlake.edu.cn

**Abstract**
Semimetals exhibit intriguing characteristics attributed to the coexistence of both electrons and holes. In rhombohedral multilayer graphene, a strong trigonal warping effect gives rise to a semi-metallic state near the Fermi surface, offering unique opportunities to explore the interplay of semi-metallic properties with strong correlations and topologies. Here, the observation of quarter semimetals in rhombohedral multilayer graphene by introducing spin-orbit coupling (SOC) is reported. The semi-metallic characteristics of rhombohedral graphene manifest as nearly vanished Hall resistance and parabolic longitudinal resistance. The strong correlations arising from the surface flat band lead to spontaneous symmetry breaking. SOC proximitized by $WSe_2$ further lifts the valley degeneracy, resulting in the spontaneous time-reversal symmetry breaking, as evidenced by the hysteretic anomalous Hall effect. The coexistence of fully polarized electrons and holes allows for the observation of a non-monotonic temperature dependence of the anomalous Hall resistance. Furthermore, the application of moderate magnetic fields induces a phase transition from quarter semimetals to Chern insulators. These findings establish rhombohedral multilayer graphene as an ideal platform for studying strong correlations and topologies in semimetals.

A semimetal is characterized by a small overlap in the energy of its conduction and valence bands, resulting in the coexistence of electrons and holes near the Fermi surface. This unique electronic structure makes semimetals an ideal platform for exploring various exotic quantum phenomena and designing advanced electronic devices. For example, giant magnetoresistance has been reported in compensated semimetals such as $WTe_2$, bismuth, NbP, and graphene[1-4]. Additionally, the strong Coulomb attraction between electrons and holes in semimetals can lead to the formation of excitonic insulators, which support neutral bosons and are considered as potential candidates for high-temperature Bose-Einstein condensates and excitonic superfluids[5-7]. In graphene semimetal, the interacting electron-hole plasma, known as the Dirac fluid, exhibits hydrodynamic behavior, leading to the violation of the Wiedemann-Franz law and the observation of quantum critical conductivity[8-10]. Topological semimetals, such as Dirac, Weyl, and nodal-line semimetals, host a variety of topologically nontrivial electronic states[11,12]. Breaking time-reversal symmetry in semimetals will stimulate the emergence of even more exotic states and has potential applications in spintronics, yet suitable materials are still rare[13-16].

Rhombohedral multilayer graphene has recently emerged as a novel platform to explore the interplay between strong correlations and topologies, owing to its layer-dependent surface flat band and large momentum-space Berry curvatures near the Fermi energy[17-19]. Consequently, correlated many-body gaps have been demonstrated in rhombohedral trilayer and tetralayer graphene[20-22]. When placed on h-BN with high-quality interface, rhombohedral multilayer graphene shows versatile interaction-driven spontaneous symmetry breaking states[23-29]. Rhombohedral trilayer graphene and tetralayer graphene exhibit intricate superconductivity with unconventional pairings[30,31]. By further introducing spin-orbit coupling (SOC) in rhombohedral graphene in contact with transition metal dichalcogenides (TMDC), high-Chern-number Chern insulators and quantum anomalous Hall insulators have been observed in tetralayer and pentalayer, respectively[32,33]. Even more remarkably, rhombohedral graphene/h-BN moiré superlattices have been shown to host correlated Chern bands and fractional quantum anomalous Hall effect[34-37].

The flat conduction and valence bands of rhombohedral multilayer graphene can be approximately characterized by a simple two-band model, producing a power-law energy dispersion. With the increasing layer numbers, high order hopping terms strikingly distort the surface flat band by means of trigonal warping effect, resulting in a semi-metallic band structure near the Fermi surface[18,19]. The overlap of the conduction and valence bands in semi-metallic rhombohedral graphene can be precisely controlled by external electric field[18,24]. Thus, rhombohedral multilayer graphene serves as an ideal platform for investigating time-reversal symmetry broken semimetals, which is still to be explored.

In this work, we report the observation of quarter semimetals in rhombohedral pentalayer graphene, achieved by introducing SOC through the proximity effect. This observation is a result of the interplay between correlations, topology, and SOC. The coexistence of electrons and holes is evidenced by nearly vanished Hall resistance and quadratic dependence of longitudinal resistance at low magnetic fields, which can be well fitted by a two-carrier model. The spin-valley-locking effect lifts the valley degeneracy and breaks the time-reversal symmetry, leading to the emergence of ferromagnetism. We observed a non-monotonical temperature dependence in the anomalous Hall hysteretic loops. When moderate magnetic fields are applied, the valley-polarized semimetals experience band inversion, opening a topological nontrivial gap, which enables the observation of a high-order Chern insulator.

**Quarter-metallic states**

Our typical device (device D1) is schematically shown in Figure 1a. The proximate SOC in rhombohedral pentalayer graphene is introduced by placing a bilayer $WSe_2$ flake onto it. Previous calculations and experimental data suggest $WSe_2$ can endow graphene with an Ising SOC on the meV scale, which is several hundred times stronger than the intrinsic SOC in graphene[38-40]. To probe the subtle features within the SOC-modified band structure, we used a hexagonal boron nitride (h-BN) encapsulated graphene structure to reduce charge inhomogeneity. Raman mapping was employed to identify the preservation of rhombohedral domains after h-BN encapsulation (see Figure 1b). The dual-gate structures enable simultaneously independent control over the carrier density ($n$) and the displacement electric field ($D$).

Figure 1c shows the $n - D$ map of longitudinal resistance $R_{xx}$ in rhombohedral pentalayer graphene with SOC at $T = 0.3$ K and zero magnetic field. The phase diagram resembles that of intrinsic pentalayer graphene[26,29]. At the charge-neutrality points (CNP), there are three insulating states separated by two metallic states, namely, one layer-antiferromagnetic (LAF) insulator at small $D$ and two layer-polarized insulators (LPI) at large $D$. Due to the large density of states at the surface flat band, strong electronic interactions lead to insulating ground states at $n = 0$ and $D = 0$, which have been identified as the LAF state in previous studies[25-27]. As shown in Figure 1d, in the LAF state, electrons with opposite spins are polarized to the bottom and top layers, while maintaining their four-fold degeneracy. A small $D$ (0.1 V nm$^{-1}$ < $|D|$ < 0.25 V nm$^{-1}$) breaks the layer degeneracy and polarizes the low-energy electronic wavefunctions to either top or bottom, depending on the sign of $D$ and the layer. Consequently, two spin-valley flavors exhibit a semi-metallic band structure with an overlap of the conduction and valence bands. When $WSe_2$ is placed onto one of the surface layers, graphene strongly experiences the proximate Ising SOC, resulting in relatively large spin-splitting of the low-energy band at K and K' valley. The spin-valley locking effect of the Ising SOC causes the energy band in K (K') valley flavor to move up (down) in the valence band, further lifting the two-fold degeneracy. Therefore, at the CNP, the Fermi level crosses only one copy of the four-fold spin-valley flavors, achieving a spin-valley polarized (quarter) semimetal.

To experimentally confirm that SOC favors the emergence of quarter-metallic state near the CNP, we examined the quantum oscillations[23]. Since the quantum oscillations at CNP are weak, we slightly doped the semimetal to metallic states by tuning $n$. Figure 1e and 1f show $R_{xx}$ as a function of filling factor $\nu = nh/eB$ in the low field regime, where $h$ is the Planck constant, $e$ is the elementary charge, and $B$ is the magnetic field. At $n = -4.7 \times 10^{12}$ cm$^{-2}$, $D = 0$ V nm$^{-1}$ (Position C), a four-fold degeneracy associated with a spin-valley degenerate normal metal is observed. The degeneracy is lifted to one fold at $n = -1 \times 10^{12}$ cm$^{-2}$, $D = 0.19$ V nm$^{-1}$ (Position B) when a moderate $D$ is applied, indicating a spin-valley polarized (quarter) metal. This observation is consistent with the Fermi level labelled by $E_{F,B}$ schematized in Figure 1d. It's noted that the observed quarter metal in Position B differs from that in the spontaneously broken-symmetry metals near LPI[25]. In our case, the quarter metal appears in the immediate semimetal regions between LAF and LPI states. Furthermore, it's SOC, rather than correlations, that drives the emergence of quarter metals and semimetals.

After quarter-metallic states are already established, we further tune the Fermi level to the CNP. Figure 1g shows $R_{xy}$ and the corresponding $R_{xx}$ at $n = -1 \times 10^{11}$ cm$^{-2}$, $D = 0.17$ V nm$^{-1}$ (Position A). Unlike the linear Hall resistance $R_{xy}$ observed in hole-doped metallic states, the $R_{xy}$ at CNP exhibits remarkably nonlinear behavior. In the region of $-0.5$ T $< B < 0.5$ T, $R_{xy}$ exhibits a zero-Hall-plateau-like feature. More precisely, $R_{xy}$ in Figure 1g demonstrates multiple sign changes as a function of $B$, along with a hysteretic anomalous Hall effect at low fields of $-30$ mT $< B < 30$ mT. This behavior is quite different from the large remanent $R_{xy}$ near $B = 0$ in a SOC-induced Chern insulator recently observed in rhombohedral multilayer graphene, where a topologically nontrivial gap is opened due to large SOC-induced band inversion[32,33]. In our system, the SOC is strong enough to lift the valley degeneracy, but weak enough to prevent the gap opening, thus maintaining the semimetal characteristics, consistent with the theoretical calculations[41]. In the following sections, we will demonstrate that the observed nonlinearity and anomalous Hall effect arise from the time-reversal symmetry broken semimetals in our system.

**Two-carrier model**

To elucidate the nonlinearity and nearly vanished $R_{xy}$ at low magnetic fields, we performed a temperature evolution study of $R_{xx}$ and $R_{xy}$ and employed the two-carrier model to describe the transport behavior near the semi-metallic states. In a semimetal, both electrons and holes contribute to the conductance. The longitudinal and Hall resistivity under magnetic fields can be described by the following expressions[42]:

$$\rho_{xx}(B) = \frac{(n_e\mu_e + n_h\mu_h) + (n_e\mu_e\mu_h^2 + n_h\mu_h\mu_e^2)B^2}{e[(n_e\mu_e + n_h\mu_h)^2 + (n_h - n_e)^2 \mu_e^2 \mu_h^2 B^2]} \quad (1)$$

$$\rho_{xy}(B) = \frac{(n_h\mu_h^2 - n_e\mu_e^2)B + \mu_h^2\mu_e^2(n_h - n_e)B^3}{e[(n_e\mu_e + n_h\mu_h)^2 + (n_h - n_e)^2 \mu_e^2 \mu_h^2 B^2]} \quad (2),$$

where $n_e$, $n_h$, $\mu_e$, and $\mu_h$ are the densities of electron and hole, and the carrier mobilities of electron and hole, respectively. The multiband conduction can give rise to a nonlinear Hall effect, which has been widely observed in various systems[43-45]. In the case of an ideal compensated semimetal where $n_e = n_h$ and the mobilities of electrons and holes are equal ($\mu_e = \mu_h$), the system will exhibit parabolic dependences of $\rho_{xx}(B)$ and vanished $\rho_{xy}(B)$ at small $B$.

As shown in Figure 2, we used Equations (1) (2) to quantitatively fit the measured $R_{xx}$ and $R_{xy}$ at various temperatures for a fixed $n = -1 \times 10^{11}$ cm$^{-2}$, $D = 0.17$ V nm$^{-1}$ (Position A in Figure 1c). To this end, we subtracted the anomalous Hall components arising from time-reversal symmetry breaking, which we will discuss in detail later. The detailed fitting procedure is described in the Supplementary Information. The remaining nonlinear components of $R_{xy}$, along with their corresponding $R_{xx}$, can be well fitted by the aforementioned two-carrier model across all temperature regions. Figure 2c shows the comparison of experimental data and the fitted curves at a representative temperature ($T = 0.3$ K). The fitting results yield nearly linear $R_{xy}$ at high temperatures. Pronounced nonlinear $R_{xy}$ emerges at $T < 5$ K and is significantly enhanced as the temperatures decreases. The Hall coefficient, defined as $dR_{xy}/dB$, changes its sign at $T = 1$ K at low field ($B \approx 0$ T) as the temperature decreases. At fixed $T = 0.3$ K, the Hall coefficient also experiences sign reversal at $B \approx 0.3$ T. The observations of complex nonlinearity in $\rho_{xy}(B)$ and nearly parabolic dependences of $\rho_{xx}(B)$ agree well with the two-carrier model, which allows us to unambiguously confirm the semimetallic nature of rhombohedral graphene[42-45].

We further extracted electron and hole densities, as well as their mobilities, which are plotted as a function of temperature in Figure 2e and 2f. Both $\mu_e$ and $\mu_h$ increase with decreasing temperature, reaching $10^4$ cm$^2$ V$^{-1}$s$^{-1}$ at low temperatures. Compared with monolayer graphene with linear energy dispersion[3], the carrier mobilities in rhombohedral graphene semimetals are relatively lower, originating from their low-energy flat band characteristic. Meantime, both $n_e$ and $n_h$ decrease with decreasing temperature. The values of $n_e$ and $n_h$ at $T = 0.3$ K are in the order of $10^{11}$ cm$^{-2}$, indicating small hole and electron pockets near CNP arising from the fine structure of trigonally warped rhombohedral graphene[18,19]. The observed temperature evolution of carrier densities and mobilities is consistent with the picture that both electrons and holes are thermally activated and scattered by phonons at high temperatures. This feature further confirms the coexistence of electrons and holes in our system.

**Hysteretic anomalous Hall effect**
Besides the nonlinearity, we observe hysteretic loops in $R_{xy}$ at low-field regimes ($|B| < 50$ mT) when sweeping $B$ forward and backward. This phenomenon arises from the interplay of SOC, strong correlations, and topology in our system. As illustrated in Figure 1d, proximate SOC, in conjunction with strong correlations in rhombohedral multilayer graphene, produces valley-polarized semimetals. In this scenario, both electrons and holes are valley-polarized with a large Berry curvature, endowing them with a giant orbital magnetic moment[18,46]. Since the K and K' valleys in graphene are related by time-reversal symmetry, lifting the valley degeneracy spontaneously induces time-reversal symmetry breaking. Consequently, we observe ferromagnetic states, manifesting as the anomalous Hall effect with hysteretic loops at low-field regime.

Figure 3 depicts the temperature evolution of the hysteretic anomalous Hall effect. Intriguingly, the remanent Hall resistance $\Delta R_{xy}$ (defined in Methods), which is proportional to the magnetization, exhibits a non-monotonical temperature dependence. In the temperature range of 5 K $< T <$ 10 K, $\Delta R_{xy}$ increases with decreasing temperature, which is a common observation in graphene-based ferromagnetic systems[35,47-49]. However, with further decreasing temperature, $\Delta R_{xy}$ dramatically decreases. We attribute this behavior to a competition between two distinct effects governing net magnetization. The first is thermal suppression of magnetic order. As in conventional ferromagnets, increasing temperature provides energy for thermal spin fluctuations, which disrupts the spontaneous magnetic order and tends to decrease the net magnetization. The second is temperature-dependent magnetic moments density. The magnetism in our system is mediated by two types of itinerant magnetic moments (valley-polarized electrons and holes dressed by orbital moments). As shown in Figure 2e, the density of these magnetic carriers is itself highly temperature-dependent. As temperature decreases, this magnetic moment density decreases, leading to a reduction of the net magnetization. These two effects have opposing influences as a function of temperature. The peak in the anomalous Hall effect occurs at the moderate temperature where the suppression of thermal fluctuations (which increases magnetization as $T$ is lowered) is balanced by the reduction in magnetic carrier density (which decreases magnetization as $T$ is lowered).

We can exclude the domain wall effects leading to abnormal temperature dependence by repetitively measuring the temperature dependence in several cooling down processes. If domain wall effects dominate the mechanism, when the sample being warmed up above Curie temperature, the magnetic domains randomize and rearrange after cooling down again. However, we observed consistent temperature dependence behaviors across different cooling processes, as shown in Figure 3b. Our observation of non-monotonic temperature dependence of $\Delta R_{xy}$ highlights the intricate interplay between thermal effects and the magnetic properties of the system, underscoring the complex nature of ferromagnetism in rhombohedral multilayer graphene.

It's worth noting that although there is no moiré superlattice in our system, we observe multiple Barkhausen jumps upon sweeping $B$ in the hysteresis loops, indicating the presence of multiple magnetic domains in this system[48,50]. Notably, there is a special magnetic domain structure in our system. The valley-polarized electrons and holes with different coercivities due to their distinct magnetic moments and mobilities can contribute to the domain structures induced Barkhausen jumps.

**Topological phase transition to Chern insulators**

The spin-valley locking effect of the Ising SOC allows for magnetic control of valleys through the valley Zeeman effect[51-53]. The applied external magnetic field directly couples to the orbital magnetic moments of the Bloch electrons. This coupling is valley-dependent due to the opposite Berry curvature in the K and K' valley. The valley Zeeman field further breaks the valley degeneracy by enhancing the band overlap and eventually inverting the band in the valley-polarized semi-metallic band, while enlarging the gap in the other valley, as shown in Figure 4a. By fully polarizing the system into a single valley, the field stabilizes a state where the exchange interaction-driven band gap is maximized. The system transitions from a semimetal (where the Fermi level crosses a band) to a true insulator (where the Fermi level lies within a gap), as shown in Figure 4a. Crucially, the occupied band below this gap has a non-zero Chern number, transforming the system into a Chern insulator[47,54-56].

As shown in Figure 4, under sufficiently large $B$, we observe the topological phase transition from the valley-polarized semimetals to Chern insulators. In Figure 4d, at the base temperature of 0.3 K, starting from $B \approx 0.8$ T, $|R_{xy}|$ dramatically increases and quantizes to $h/Ce^2$ at $B \approx 1.5$ T, where $C$ is the Chern number. We find $C = -5$ here, precisely matching the winding number of rhombohedral pentalayer graphene[17,25,26,57]. The sign of $R_{xy}$ can be reversal by changing the sign of $B$. Further increasing $B$ brings the system to the coexistence of Chern states and quantum hall states, resulting in the decrease of $R_{xy}$ and the quantization at higher Chern number. The Chern insulators can be further confirmed by tracing $R_{xy}$ and $R_{xx}$ with the evolution of $B$ and $n$ in the fan diagram shown in Figure 4b and 4c. The positions of minimum $R_{xx}$ in Figure 4b and large $R_{xy}$ with quantized values in Figure 4c can be well described by the Streda's formula $\frac{\partial n}{\partial B} = C\frac{e}{h}$. From the slope of the dashed lines in Figure 4b and 4c, again we obtained $C = -5$, consistent with the quantized value in Figure 4d. It is worth noting that our observations of Chern insulators at moderate $B$ differ from previous zero-field Chern insulators[25,26], because the ground state of our system is a quarter semimetal, while the latter are topologically gapped states.

In the Chern states regime ($B = 1.5$ T), as shown in Figure 4e, the temperature dependence of $R_{xy}$ exhibits a normal monotonic behavior, namely, $R_{xy}$ increases with decreasing temperature and saturates at ultra-low temperatures. This behavior contrasts sharply with that observed in the semi-metallic regime, as shown in Figure 3b, where $\Delta R_{xy}$ demonstrates non-monotonic temperature dependence. The distinct monotonic feature in the Chern states regime provides strong evidence for the opening of topological gaps and the phase transition occurring at moderate $B$.

**Conclusion**

Our findings establish SOC proximitized rhombohedral multilayer graphene as unique ferromagnetic semimetals, which originate from the trigonally warped band structure, strong correlations, and SOC induced spontaneous time-reversal symmetry breaking. The intertwin of strong correlations, SOC, and topology in this system offers a highly tunable platform for exploring a wealth of emergent phenomena. Compared with recently discovered zero-field quantum anomalous Hall insulators in $WS_2$/rhombohedral pentalayer graphene[33], our results underscore the crucial role of SOC strength in determining various quantum states. Theoretical studies suggest that the strength of effective proximity-induced SOC strongly depends on the rotational alignment between graphene and $WSe_2$. The strongest SOC effects are predicted to occur when graphene is rotationally misaligned with $WSe_2$ by approximately 15° to 20°[58]. We have identified the twist angles between $WSe_2$ and graphene in this study to be ~18° for device D1 and ~15° for device D2 (see Supplementary Figure S2). These angles fall within the optimal range predicted to induce the strongest SOC in graphene/$WSe_2$ structure. Therefore, we believe that the material selection ($WSe_2$ rather than $WS_2$) involving a complex interplay of intrinsic strength and interlayer coupling is a more decisive factor for observing the quarter-semimetal state in our devices, when compared to the quantum anomalous Hall state previously reported[33]. Our work motivates future efforts to systematically investigate the evolution of the phase diagram with the SOC strength, which can be achieved by either designing special structures to control the twist angle between graphene and TMDC[40,58,59], or tuning the interlayer coupling strength through hydrostatic pressure[60]. Furthermore, proximate anisotropic SOC on rhombohedral graphene is worth studying by using low-symmetry two-dimensional materials with large anisotropic SOC.

Moreover, the flat band along with the coexistence of electrons and holes in rhombohedral graphene semimetals enables the studies of strongly correlated electron-hole states[8,61], such as excitonic insulators with superfluid states and viscous Dirac fluids with hydrodynamic transport. Combined with the topological features of rhombohedral graphene, our system is also a suitable candidate for exploring the fractional quantum anomalous Hall effect in a semimetal[62,63].

**Methods**

**Device fabrication**

Single crystals of WSe$_2$ were grown by the chemical vapor transport method. Graphite crystals were commercially purchased from NGS Naturgraphit. Multilayer graphene and WSe$_2$ flakes were mechanically exfoliated from bulk crystals onto SiO$_2$/Si substrates. The stacking orders of multilayer graphene were identified by Raman spectroscopy (WITec alpha300). The layer number of graphene was initially identified by optical contrast using an optical microscope (Nikon LV100ND) and subsequently confirmed by the transport measurements. To reduce the possibility of domain relaxations during the transfer process, the rhombohedral stacking domains were isolated from Bernal stacking domains in advance by cutting the flake using a tungsten tip. To assemble the h-BN encapsulated heterostructures, we utilized the standard dry transfer method with the assistance of a polypropylene carbonate/polydimethylsiloxane (PC/PDMS) stamp. We sequentially picked up the top h-BN, WSe$_2$, and pentalayer graphene, and released them onto a pre-prepared bottom h-BN/Pt metallic bottom gate. The final heterostructure was further checked by Raman mapping to ensure that the rhombohedral stacking was alive. Atomic force microscopy was used to select the bubble-free regions. Subsequently, the heterostructure was fabricated into a device with Hall bar geometry using standard electron-beam lithography, reactive-ion etching, and electron-beam evaporation. We have fabricated three kinds of devices. The data shown in the main text were acquired from rhombohedral pentalayer graphene in contact with bilayer WSe$_2$ (device D1). We also show the data from rhombohedral pentalayer graphene in contact with monolayer WSe$_2$ (device D2) and intrinsic rhombohedral pentalayer graphene without WSe$_2$ (device D3) in Supplementary Figure S10 and Supplementary Figure S11, respectively.

**Transport measurement**

The transport measurements were performed in a cryogen-free variable temperature cryostat (Cryogenic Limited) equipped with a helium-3 insert. The base temperature was maintained at 0.3 K. Standard lock-in amplifiers (SR830 and SR860, Stanford Research Systems) were used to measure the longitudinal and Hall resistance at a frequency of 19.49 Hz. An AC current source provided 10 nA for the measurements of magnetic states and 100 nA for other measurements. Gate voltages were applied using a Keithley 2636B source meter.

The independent control over the charge carrier density $n$ and electric displacement field $D$ was achieved using our dual-gate structure. They are calculated as follows: $n = \frac{C_b \Delta V_b}{e} + \frac{C_t \Delta V_t}{e}$ and $D = \frac{C_b \Delta V_b - C_t \Delta V_t}{2\varepsilon_0}$, where $C_b$ ($C_t$) are the normalized bottom (top)-gate capacitances, $\Delta V_b = V_b - V_b^0$ ($\Delta V_t = V_t - V_t^0$) are the effective bottom (top) gate voltages, $V_b$ ($V_t$) are the applied bottom (top) gates, $e$ is the elementary charge, and $\varepsilon_0$ is the vacuum permittivity.

**Normalized FFT frequency**

The quantum oscillations were analyzed by applying Fast Fourier Transform (FFT) to the data of $R_{xx}$ as a function of $1/B$. The oscillation frequency was normalized by the total carrier density $n$, defined as $f_\nu = f_{1/B}/n\phi_0$, where $f_{1/B}$ is the frequency by deriving the quantum oscillation, $\phi_0 = h/e$ is the magnetic flux quantum, and $\nu = n\phi_0/B$ is the filling factor of Landau levels. The normalized frequency $f_v$ indicates the fraction of the total area of the Fermi surface when intersected by the Fermi surface.

**Data analysis of Hall resistance**

The measured Hall resistance $R_{xy}$ consists of two components $R_{xy} = R_{\text{NH}} + R_{\text{AH}}$, where $R_{\text{NH}}$ is attributed to the nonlinear Hall effect due to the coexistence of electrons and holes, and $R_{\text{AH}}$ arises from the anomalous Hall effect due to Berry curvature-induced anomalous velocity. To isolate these components, we first subtracted $R_{\text{AH}}$ from the measured $R_{xy}$ based on the hysteretic anomalous Hall signals, assuming the saturation magnetization (and thus the saturation $R_{\text{AH}}$) remains constant at high magnetic field. The remaining $R_{\text{NH}}$ was then fitted using a two-carrier model. A typical process can be found in Supplementary Figure S3.

Experimentally, $R_{xx}$ and $R_{xy}$ often mix with each other even when using standard Hall bar geometry. To separate them, we employed standard symmetric and anti-symmetric procedures to obtain pure $R_{xx}$ and $R_{xy}$, respectively, as follows: $\rho_{xx}(B,\rightarrow) = \frac{R_{xx}(B,\rightarrow)+R_{xx}(-B,\leftarrow)}{2}\frac{W}{L}$, $\rho_{xy}(B,\rightarrow) = \frac{R_{xy}(B,\rightarrow)-R_{xy}(-B,\leftarrow)}{2}$, where $W/L$ is the width-to-length ratio of the Hall bar, designed to be 1 during fabrication. In Figure 4b and 4c, we applied $R_{xy}(\pm B) = [R_{xy}(B) - R_{xy}(-B)]/2$ and $R_{xx}(\pm B) = [R_{xx}(B) + R_{xx}(-B)]/2$, respectively.

In Figure 2, we used the above symmetric and anti-symmetric treatments in order to fit the two-carrier model, while we reserved the raw data without any processing in other Figures.

The $\Delta R_{xy}$ shown in Figure 3b is defined as $\Delta R_{xy} = (R_{xy}(B > |B_c|) - R_{xy}(B < -|B_c|))/2$, where $R_{xy}(B > |B_c|)$ and $R_{xy}(B < -|B_c|)$ refer to the corresponding Hall resistance at saturated magnetization, $B_c$ is the coercive magnetic field of the hysteretic anomalous Hall signal.

**Two-carrier model fitting**

After obtaining the symmetric $\rho_{xx}$ and anti-symmetric $\rho_{xy}$, we fitted the carrier density and mobility of electrons and holes using the two-carrier model described in Equations (1) and (2).

To minimize the influence of the hysteretic anomalous Hall component and Chern states, we performed the fitting using data outside these regions. Specifically, at $0.3 \text{ K} \leq T \leq 4 \text{ K}$, we fitted $\rho_{NL}$ in the region of $0.1 \text{ T} \leq B \leq 0.5 \text{ T}$. When the temperature increased to $4.5 \text{ K} \leq T \leq 7 \text{ K}$, we extended $B$ range up to 1.1 T due to the smearing of Chern states. At even higher temperatures of $8 \text{ K} \leq T \leq 20 \text{ K}$, we extended $B$ range up to 2 T.

The typical data fitting process is shown in Supplementary Figure S3, and the fitting results are shown in Figure 2c. Each fitting curve yields four parameters: $n_e$, $n_h$, $\mu_e$, and $\mu_h$. The corresponding values with error bars fitted at various temperatures are shown in Figure 2e and 2f.

**References:**

1. Yang, F. Y. *et al.* Large magnetoresistance of electrodeposited single-crystal bismuth thin films. *Science* **284**, 1335-1337 (1999).
2. Ali, M. N. *et al.* Large, non-saturating magnetoresistance in WTe$_2$. *Nature* **514**, 205-208 (2014).
3. Xin, N. *et al.* Giant magnetoresistance of Dirac plasma in high-mobility graphene. *Nature* **616**, 270-274 (2023).
4. Shekhar, C. *et al.* Extremely large magnetoresistance and ultrahigh mobility in the topological Weyl semimetal candidate NbP. *Nat. Phys.* **11**, 645-649 (2015).
5. Kogar, A. *et al.* Signatures of exciton condensation in a transition metal dichalcogenide. *Science* **358**, 1314-1317 (2017).
6. Jia, Y. *et al.* Evidence for a monolayer excitonic insulator. *Nat. Phys.* **18**, 87-93 (2022).
7. Sun, B. *et al.* Evidence for equilibrium exciton condensation in monolayer WTe$_2$. *Nat. Phys.* **18**, 94-99 (2022).
8. Crossno, J. *et al.* Observation of the Dirac fluid and the breakdown of the Wiedemann-Franz law in graphene. *Science* **351**, 1058-1061 (2016).
9. Gallagher, P. *et al.* Quantum-critical conductivity of the Dirac fluid in graphene. *Science* **364**, 158-162 (2019).
10. Ku, M. J. H. *et al.* Imaging viscous flow of the Dirac fluid in graphene. *Nature* **583**, 537-541 (2020).
11. Liu, Z. K. *et al.* Discovery of a three-dimensional topological Dirac semimetal, Na$_3$Bi. *Science* **343**, 864-867 (2014).
12. Lu, L. *et al.* Experimental observation of Weyl points. *Science* **349**, 622-624 (2015).
13. Liu, E. *et al.* Giant anomalous Hall effect in a ferromagnetic kagome-lattice semimetal. *Nat. Phys.* **14**, 1125-1131 (2018).
14. Matsuoka, H. *et al.* Band-driven switching of magnetism in a van der Waals magnetic semimetal. *Sci. Adv.* **10**, eadk1415 (2024).
15. Kim, K. *et al.* Large anomalous Hall current induced by topological nodal lines in a ferromagnetic van der Waals semimetal. *Nat. Mater.* **17**, 794-799 (2018).
16. Liu, J. Y. *et al.* A magnetic topological semimetal Sr$_{1-y}$Mn$_{1-z}$Sb$_2$ (y, z < 0.1). *Nat. Mater.* **16**, 905-910 (2017).
17. Zhang, F., Jung, J., Fiete, G. A., Niu, Q. & MacDonald, A. H. Spontaneous quantum Hall states in chirally stacked few-layer graphene systems. *Phys. Rev. Lett.* **106**, 156801 (2011).
18. Slizovskiy, S., McCann, E., Koshino, M. & Fal'ko, V. I. Films of rhombohedral graphite as two-dimensional topological semimetals. *Commun. Phys.* **2**, 164 (2019).
19. Koshino, M. & McCann, E. Trigonal warping and Berry's phase N$\pi$ in ABC-stacked multilayer graphene. *Phys. Rev. B* **80**, 165409 (2009).
20. Myhro, K. *et al.* Large tunable intrinsic gap in rhombohedral-stacked tetralayer graphene at half filling. *2D Mater.* **5**, 045013 (2018).
21. Kerelsky, A. *et al.* Moiréless correlations in ABCA graphene. *Proc. Natl. Acad. Sci. U.S.A.* **118**, e2017366118 (2021).
22. Lee, Y. *et al.* Competition between spontaneous symmetry breaking and single-particle gaps in trilayer graphene. *Nat. Commun.* **5**, 5656 (2014).
23. Zhou, H. X. *et al.* Half- and quarter-metals in rhombohedral trilayer graphene. *Nature* **598**, 429-433 (2021).

24  Shi, Y. *et al.* Electronic phase separation in multilayer rhombohedral graphite. *Nature* **584**, 210-214 (2020).

25  Liu, K. *et al.* Spontaneous broken-symmetry insulator and metals in tetralayer rhombohedral graphene. *Nat. Nanotechnol.* **19**, 188-195 (2024).

26  Han, T. *et al.* Correlated insulator and Chern insulators in pentalayer rhombohedral-stacked graphene. *Nat. Nanotechnol.* **19**, 181-187 (2024).

27  Zhou, W. *et al.* Layer-polarized ferromagnetism in rhombohedral multilayer graphene. *Nat. Commun.* **15**, 2597 (2024).

28  Chen, G. *et al.* Evidence of a gate-tunable Mott insulator in a trilayer graphene moiré superlattice. *Nat. Phys.* **15**, 237-241 (2019).

29  Han, T. *et al.* Orbital multiferroicity in pentalayer rhombohedral graphene. *Nature* **623**, 41-47 (2023).

30  Zhou, H. X., Xie, T., Taniguchi, T., Watanabe, K. & Young, A. F. Superconductivity in rhombohedral trilayer graphene. *Nature* **598**, 434-438 (2021).

31  Han, T. *et al.* Signatures of chiral superconductivity in rhombohedral graphene. *Nature* **643**, 654–661 (2025).

32  Sha, Y. *et al.* Observation of a Chern insulator in crystalline ABCA-tetralayer graphene with spin-orbit coupling. *Science* **384**, 414-419 (2024).

33  Han, T. *et al.* Large quantum anomalous Hall effect in spin-orbit proximitized rhombohedral graphene. *Science* **384**, 647-651 (2024).

34  Lu, Z. *et al.* Fractional quantum anomalous Hall effect in multilayer graphene. *Nature* **626**, 759-764 (2024).

35  Chen, G. *et al.* Tunable correlated Chern insulator and ferromagnetism in a moiré superlattice. *Nature* **579**, 56-61 (2020).

36  Han, X. *et al.* Engineering the band topology in a rhombohedral trilayer graphene moiré superlattice. *Nano Lett.* **24**, 6286-6295 (2024).

37  Ding, J. *et al.* Electric-field switchable chirality in rhombohedral graphene Chern insulators stabilized by tungsten diselenide. *Phys. Rev. X* **15**, 011052 (2025).

38  Gmitra, M. & Fabian, J. Proximity effects in bilayer graphene on monolayer $WSe_2$: field-effect spin valley locking, spin-orbit valve, and spin transistor. *Phys. Rev. Lett.* **119**, 146401 (2017).

39  Island, J. O. *et al.* Spin-orbit-driven band inversion in bilayer graphene by the van der Waals proximity effect. *Nature* **571**, 85-89 (2019).

40  Zhang, Y. *et al.* Enhanced superconductivity in spin–orbit proximitized bilayer graphene. *Nature* **613**, 268-273 (2023).

41  Liu, Z. & Wang, J. Layer-dependent quantum anomalous Hall effect in rhombohedral graphene. *Phys. Rev. B* **111**, L081111 (2025).

42  Sondheimer, E. H. & Wilson, A. H. The theory of the magneto-resistance effects in metals. *Proc. R. Soc. Lond. A* **190**, 435-455 (1947).

43  Wang, L. *et al.* Tuning magnetotransport in a compensated semimetal at the atomic scale. *Nat. Commun.* **6**, 8892 (2015).

44  Liu, Y. *et al.* Gate-tunable multiband transport in $ZrTe_5$ thin devices. *Nano Lett.* **23**, 5334-5341 (2023).

45  Ding, D. *et al.* Multivalley superconductivity in monolayer transition metal dichalcogenides. *Nano Lett.* **22**, 7919-7926 (2022).

46  Chen, G. R. *et al.* Tunable orbital ferromagnetism at noninteger filling of a moire superlattice. *Nano Lett.* **22**, 238-245 (2022).

47  Serlin, M. *et al.* Intrinsic quantized anomalous Hall effect in a moiré heterostructure. *Science* **367**, 900-903 (2020).

48  Sharpe, A. L. *et al.* Emergent ferromagnetism near three-quarters filling in twisted bilayer graphene. *Science* **365**, 605-608 (2019).

49  Polshyn, H. *et al.* Electrical switching of magnetic order in an orbital Chern insulator. *Nature* **588**, 66-70 (2020).

50  Lin, J. X. *et al.* Spin-orbit-driven ferromagnetism at half moiré filling in magic-angle twisted bilayer graphene. *Science* **375**, 437-441 (2022).

51  Srivastava, A. *et al.* Valley Zeeman effect in elementary optical excitations of monolayer $WSe_2$. *Nat. Phys.* **11**, 141-147 (2015).

52  Aivazian, G. *et al.* Magnetic control of valley pseudospin in monolayer $WSe_2$. *Nat. Phys.* **11**, 148-152 (2015).

53  Zhao, W. *et al.* Realization of the Haldane Chern insulator in a moiré lattice. *Nat. Phys.* **20**, 275-280 (2024).

54  Chang, C.-Z. *et al.* Experimental observation of the quantum anomalous Hall effect in a magnetic topological insulator. *Science* **340**, 167-170 (2013).

55  Deng, Y. *et al.* Quantum anomalous Hall effect in intrinsic magnetic topological insulator $MnBi_2Te_4$. *Science* **367**, 895-900 (2020).

56  Li, T. *et al.* Quantum anomalous Hall effect from intertwined moiré bands. *Nature* **600**, 641-646 (2021).

57  Castro Neto, A. H., Guinea, F., Peres, N. M. R., Novoselov, K. S. & Geim, A. K. The electronic properties of graphene. *Rev. Mod. Phys.* **81**, 109-162 (2009).

58  Li, Y. & Koshino, M. Twist-angle dependence of the proximity spin-orbit coupling in graphene on transition-metal dichalcogenides. *Phys. Rev. B* **99**, 075438 (2019).

59  Zhang, Y. *et al.* Twist-programmable superconductivity in spin–orbit-coupled bilayer graphene. *Nature* (2025).

60  Fülöp, B. *et al.* Boosting proximity spin–orbit coupling in graphene/$WSe_2$ heterostructures via hydrostatic pressure. *npj 2D Mater. Appl.* **5**, 82 (2021).

61  Rickhaus, P. *et al.* Correlated electron-hole state in twisted double-bilayer graphene. *Science* **373**, 1257-1260 (2021).

62  Fu, B., Zou, J.-Y., Hu, Z.-A., Wang, H.-W. & Shen, S.-Q. Quantum anomalous semimetals. *Npj Quantum Mater* **7**, 94 (2022).

63  Yang, W. *et al.* Fractional quantum anomalous Hall effect in a singular flat band. *Phys. Rev. Lett.* **134**, 196501 (2025).

**Acknowledgements**

This work was funded by National Natural Science Foundation of China (Grant No. 12274354), the Zhejiang Provincial Natural Science Foundation of China (Grant No. LR24A040003; XHD23A2001), and Westlake Education Foundation at Westlake University. We thank Chao Zhang from the Instrumentation and Service Center for Physical Sciences (ISCPS) at Westlake University for technical support in data acquisition. We also thank Westlake Center for Micro/Nano Fabrication and the Instrumentation and Service Centers for Molecular Science for facility support. K.W. and T.T. acknowledge support from the JSPS KAKENHI (Grant Numbers 21H05233 and 23H02052) and World Premier International Research Center Initiative (WPI), MEXT, Japan.

**Author Contributions**
S.X. conceived the idea and supervised the project. H.X. fabricated the devices with the assistance of J.D.. N.L. and Z.C. grew the $WSe_2$ crystals. J.D. performed the transport measurements with the assistance of H.X.. J.D. and H.X. performed Raman measurements. K.W. and T.T. grew h-BN crystals. J.D. and S.X. analyzed data and wrote the paper. All the authors contributed to the discussions.

**Figures:**

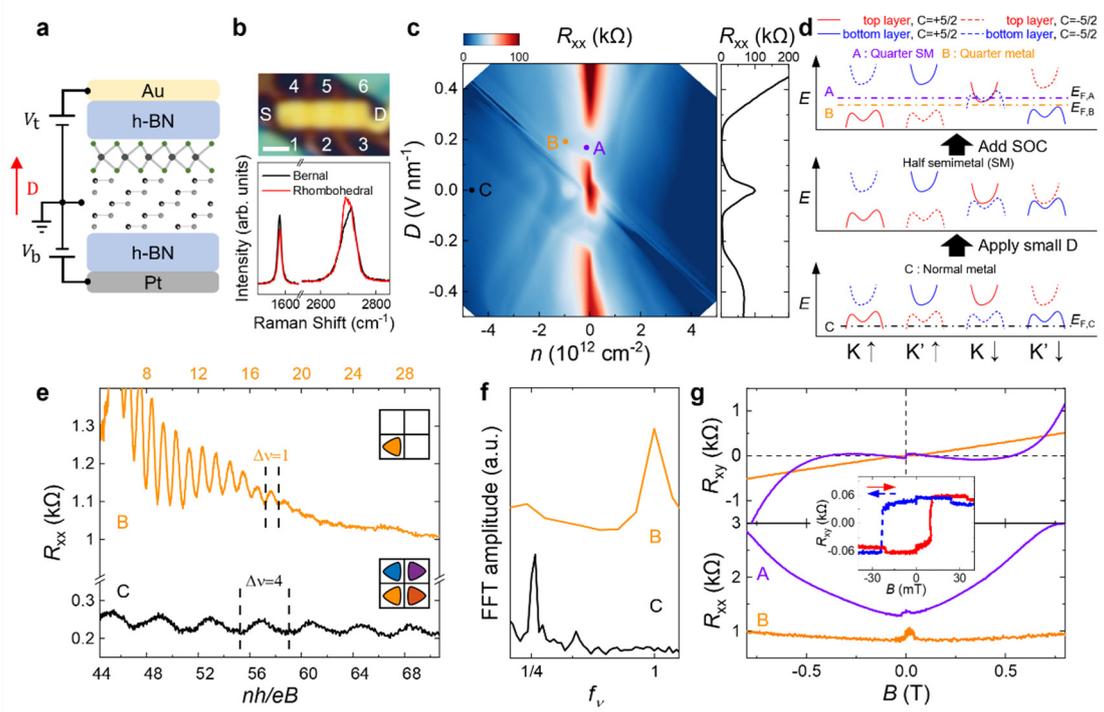

**Figure 1: Phase diagram of SOC-driven rhombohedral pentalayer graphene. a**, Schematic of the heterostructure and dual-gate configuration. **b**, Upper panel: optical image of our device. The labels mark the electrodes used to measure $R_{xx} = V_{45}/I_{SD}$ and $R_{xy} = V_{14}/I_{SD}$ in a typical configuration. The scale bar is 1 µm. Bottom panel: comparison of individual Raman spectra between rhombohedral and Bernal-stacked domains. **c**, $R_{xx}$ as a function of $n$ and $D$ at $B = 0$ T. The right curve shows $R_{xx}$ as a function of $D$ at $n = 0$ cm$^{-2}$. The labels: Position A ($n = -1 \times 10^{11}$ cm$^{-2}$, $D = 0.17$ V nm$^{-1}$), Position B ($n = -1 \times 10^{12}$ cm$^{-2}$, $D = 0.19$ V nm$^{-1}$), Position C ($n = -4.7 \times 10^{12}$ cm$^{-2}$, $D = 0$ V nm$^{-1}$). **d**, Schematic of the evolution of band structures as $D$ and SOC. The Fermi level $E_{F,A}$, $E_{F,B}$, and $E_{F,C}$ illustrates the case for Position A, B, and C marked in (**c**), respectively. The colors and curve styles distinguish the layer and valley Chern numbers, respectively. **e**, Quantum oscillations near quarter metals (position B) and normal metals (position C). The $\Delta \nu$ indicates their spin-valley degeneracies. The insets illustrate the corresponding Fermi contours. **f**, Normalized Fast Fourier Transform of the quantum oscillations in (**e**). **g**, $R_{xy}$ and $R_{xx}$ as a function of magnetic field $B$ at Position A and Position B. The inset shows the low-field magnetic hysteresis at Position A by sweeping $B$ forward and backward. All the data were measured at $T = 0.3$ K.

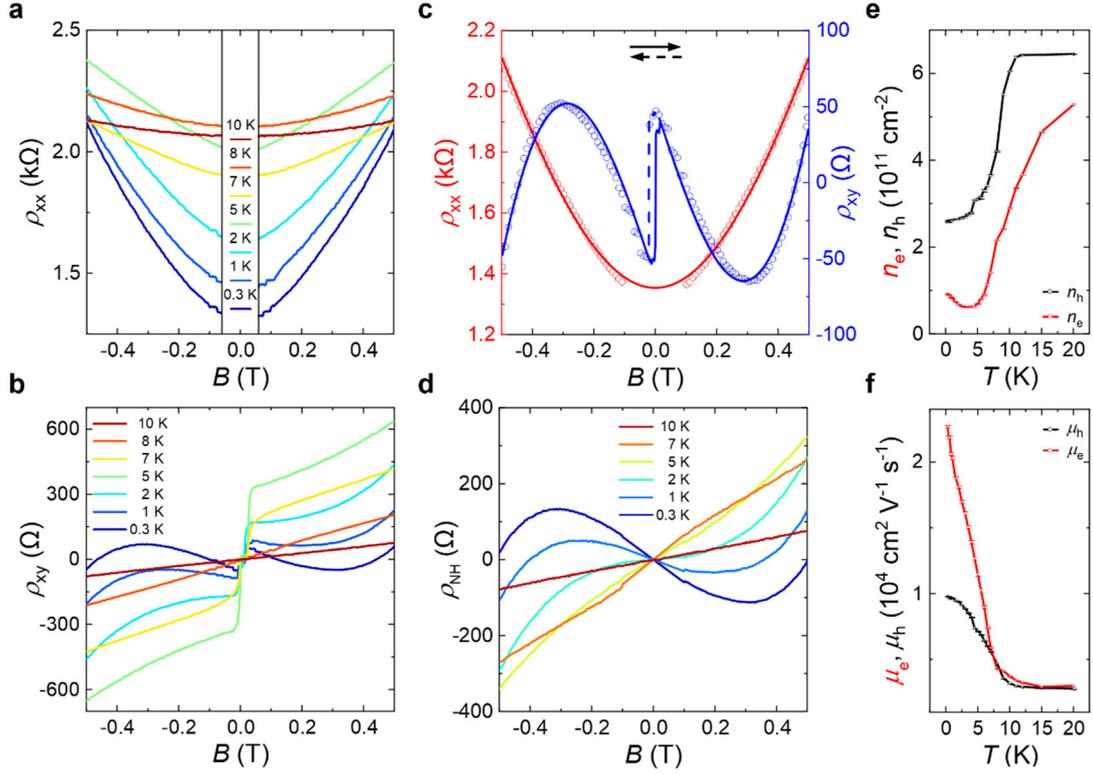

**Figure 2: Nonlinear Hall effect of semi-metallic states with the coexistence of electrons and holes. a**, Temperature dependence of longitudinal resistivity $\rho_{xx}$ as a function of $B$. **b**, Measured Hall resistivity $\rho_{xy}$ as a function of $B$ at various temperatures. The raw data at low field region in $\rho_{xx}$ were not used to fit due to its magnetic state. **c**, The comparison of the raw data (hollow symbols) and the fitted results (lines). **d**, Extracted nonlinear Hall resistivity $\rho_{NH}$ as a function of $B$ at various temperatures. **e**, Extracted electron ($n_e$, red curve) and hole ($n_h$, black curve) carrier densities. **f**, Extracted electron ($\mu_e$, red curve) and hole ($\mu_h$, black curve) carrier mobilities. The error bars were obtained from two-carrier model fitting. All the data were measured at $n = -1 \times 10^{11}$ cm$^{-2}$, $D = 0.17$ V nm$^{-1}$ marked by Position A in Figure 1c.

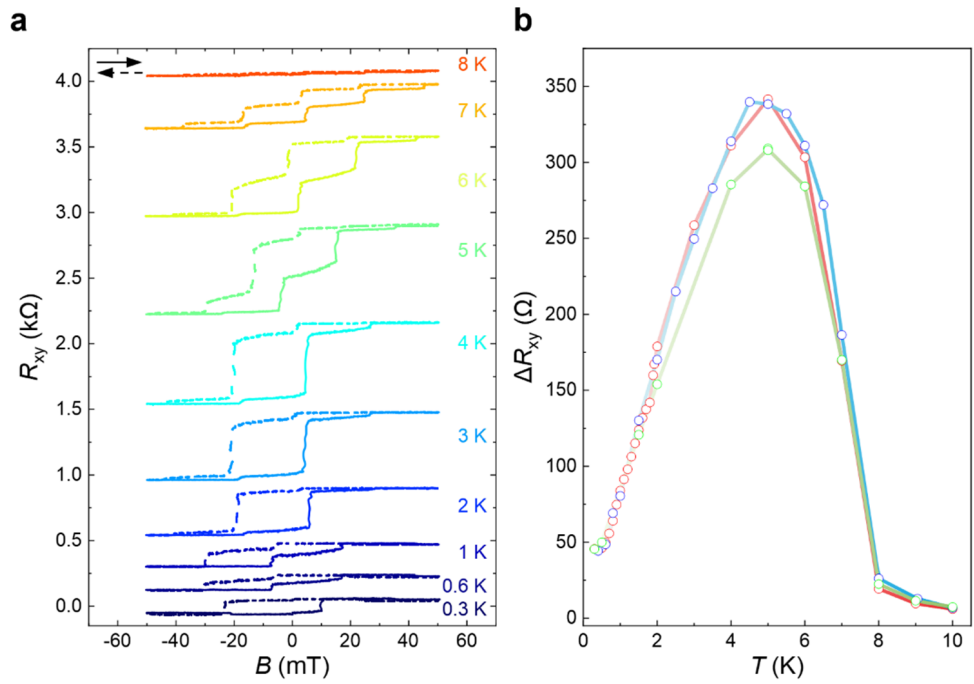

**Figure 3: Temperature dependence of hysteretic anomalous Hall effect. a**, Temperature dependence of Hall resistance $R_{xy}$ as a function of $B$ by sweeping $B$ forward (solid lines) and backward (dashed lines). **b**, The difference between forward and backward saturation resistances $\Delta R_{xy}$ as a function of temperature. Three sets of data distinguished by different colors were acquired in three different cooling down processes. All the data were measured at Position A marked in Figure 1c.

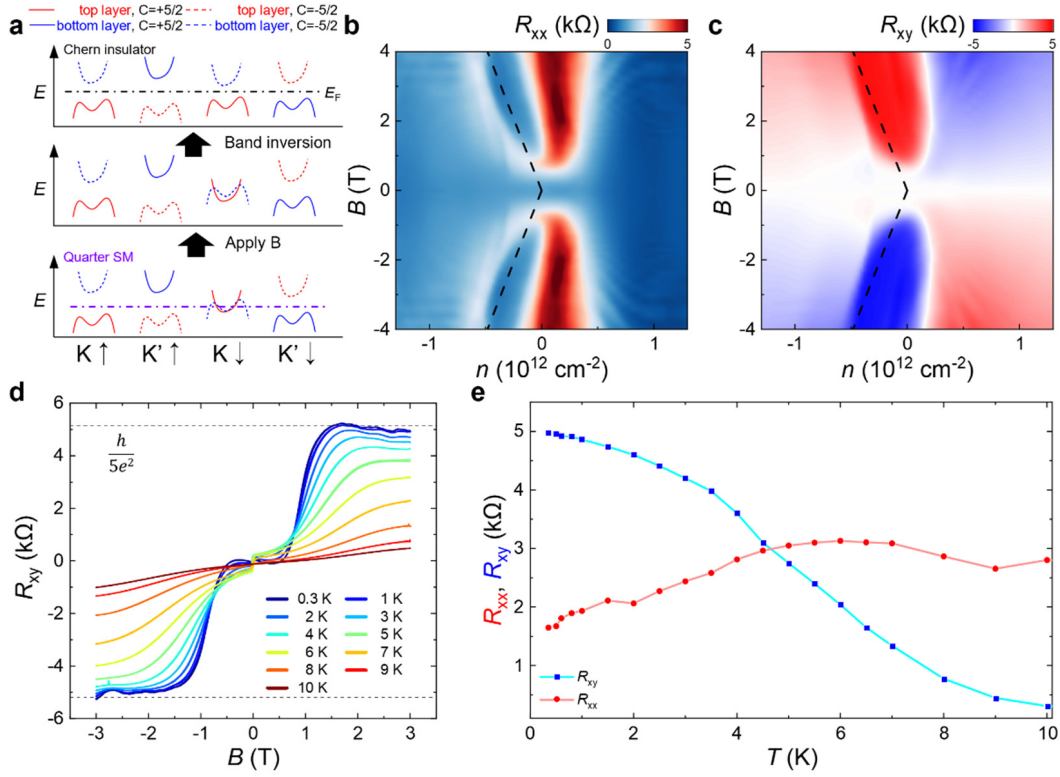

**Figure 4: Topological phase transition under magnetic fields. a**, Schematic of the evolution of band structures under magnetic fields. **b**, **c**, Symmetrized $R_{xx}$ (**b**) and anti-symmetrized $R_{xy}$ (**c**) as a function of $n$ and $B$ at a fixed $D = 0.17$ V nm$^{-1}$ and $T = 0.3$ K. The dashed lines mark the Chern states according to the Streda's formula. **d**, Temperature dependence of $R_{xy}$ (without anti-symmetry) as a function of $B$ swept at a large range between -3 T and 3 T. The dashed horizontal lines mark the Chern states quantized at $h/5e^2$. **e**, $R_{xx}$ (red dots) and $R_{xy}$ (blue dots) as a function of temperature at a fixed $B = 1.5$ T.